\documentclass[conference]{IEEEtran}
\usepackage{amsmath,amsfonts}
\usepackage{graphicx,amssymb}
\usepackage{algorithmic,algorithm}
\usepackage{amsthm,dsfont}
\usepackage{subfigure,url}
\usepackage[noadjust]{cite}
\usepackage{subfigure}

\theoremstyle{plain}
\newtheorem{lem}{Lemma}

\newcounter{longequ}[longequ]
\addtocounter{longequ}{8}

\allowdisplaybreaks

\newcommand{\Nt}{{N_\mathrm{t}}}
\IEEEoverridecommandlockouts
\begin{document}
%
\title{MISO Wireless Communication Systems via Intelligent Reflecting Surfaces }
\author{\IEEEauthorblockN{Xianghao Yu, Dongfang Xu, and Robert Schober}
\IEEEauthorblockA{Friedrich-Alexander-Universit\"{a}t Erlangen-N\"{u}rnberg, Germany\\
Email: \{xianghao.yu, dongfang.xu, robert.schober\}@fau.de}
\thanks{The work of X. Yu was supported by the Alexander von Humboldt Foundation.}
}


%

\IEEEspecialpapernotice{(Invited Paper)}

\maketitle

\begin{abstract}

Intelligent reflecting surfaces (IRSs) have received considerable attention from the wireless communications research community recently. In particular, as low-cost passive devices, IRSs enable the control of the wireless propagation environment, which is not possible in conventional wireless networks. To take full advantage of such IRS-assisted communication systems, both the beamformer at the access point  (AP) and the phase shifts at the IRS need to be optimally designed. However, thus far, the optimal design is not well understood. In this paper, a point-to-point IRS-assisted multiple-input single-output (MISO) communication system is investigated. 
The beamformer at the AP and the IRS phase shifts are jointly optimized to maximize the spectral efficiency.
Two efficient algorithms exploiting fixed point iteration and manifold optimization techniques, respectively, are developed for solving the resulting non-convex optimization problem. The proposed algorithms not only achieve a higher spectral efficiency but also lead to a lower computational complexity than the state-of-the-art approach. Simulation results reveal that deploying large-scale IRSs in wireless systems is more efficient than increasing the antenna array size at the AP for enhancing both  the spectral and the energy efficiency.

\end{abstract}

\IEEEpeerreviewmaketitle

\section{Introduction}
The capacity of current wireless networks has to exponentially increase to meet the rapidly growing demands for high-data-rate multimedia access. The network capacity can be boosted by deploying large-scale antenna arrays, network densification with small cells, and uplifting the carrier frequency to the millimeter-wave (mm-wave) bands \cite{6824752,7463025}. However, additional cost and power consumption are inevitably incurred by deploying more antenna elements, access points (APs), and radio frequency (RF) chains at extremely high frequencies (EHF) \cite{6736745}. Therefore, new paradigms that are both spectral- and energy-efficient are needed for the design of future wireless communication systems.

Recent advances in electromagnetic (EM) metasurfaces enable the manipulation of the impinging EM waves, which creates the possibility of controlling the propagation behavior of EM waves \cite{8466374}. Intelligent reflecting surfaces (IRSs), as a kind of passive metasurfaces, have been incorporated in wireless communications systems in recent years \cite{di2019smart}. In particular, IRSs are able to change the signal transmission direction with low-cost \emph{passive} devices, e.g., printed dipoles and phase shifters, which is a revolutionary new characteristic that is not  leveraged in conventional wireless communications systems. In this way, IRSs are able to create favorable wireless propagation environments while avoiding the deployment of additional energy-hungry RF chains. Furthermore, IRSs can be readily coated on the facades of  buildings, which reduces implementation cost and
complexity. However, to fully exploit the potential of energy-efficient IRSs, they have to be properly designed and integrated with conventional communication techniques, such as the transmit beamforming at APs.

There are a few previous studies on the design of IRS-assisted wireless systems \cite{nadeem2019large,8647620,wu2018intelligent,han2018large}. The IRSs are typically implemented by phase shifters that can only change the phases of the signals. A major obstacle in optimizing the phase shifts at the IRS are the associated highly non-convex unit modulus constraints. Multiple-input multiple-output (MIMO) beamforming was investigated in \cite{nadeem2019large} for IRS-assisted systems, where the phase shifts were either given or designed only for rank-one channels. A semidefinite relaxation (SDR) approach was adopted to tackle the unit modulus constraints in \cite{8647620,wu2018intelligent}. However, this approach  leads to only an approximate solution without  guarantee of optimality. In \cite{han2018large}, the design of the IRS phase shifts was simplified to a phase extraction problem based on an approximation of the ergodic capacity. In summary, despite the promising initial works \cite{nadeem2019large,8647620,wu2018intelligent,han2018large}, the general beamformer and phase shift design problem for IRS-assisted systems has not been satisfactorily solved, yet.

In this paper, we consider a point-to-point multiple-input single-output (MISO) communication system which is supported by an IRS implemented by programmable phase shifters. Two novel algorithms are proposed to jointly optimize the beamformer at the AP and the phase shifts at the IRS. 
In particular, the proposed algorithms tackle the unit modulus constraints by resorting to fixed point iteration and manifold optimization techniques, respectively. Unlike the existing results in \cite{8647620,han2018large}, both proposed algorithms guarantee locally optimal solutions for the beamformer at the AP and the IRS phase shifts. Promisingly, our simulation results show that the proposed algorithms outperform the state-of-the-art SDR method in terms of both spectral efficiency and computational complexity.

\emph{Notations:} The imaginary unit of a complex number is denoted by $\jmath=\sqrt{-1}$. Matrices and vectors are denoted by boldface capital and lower-case letters, respectively. $\mathbb{C}^{m\times n}$ denotes the set of all $m\times n$ complex-valued
matrices. 
$\mathbf{0}_m$ is the $m$-dimensional all-zero vector. 
The $i$-th element of vector $\mathbf{a}$ is denoted by $a_i$, while 
$a_{i,j}$ is the element in the $i$-th row and $j$-th column of matrix $\mathbf{A}$.
$\mathbf{A}^*$, $\mathbf{A}^T$, and $\mathbf{A}^H$ stand for the conjugate, transpose, and conjugate transpose of matrix $\mathbf{A}$, respectively. 
The $\ell_1$- and $\ell_2$-norm of vector $\mathbf{a}$ are represented as $\left\Vert\mathbf{a}\right\Vert_1$ and $\left\Vert\mathbf{a}\right\Vert_2$, respectively.
$\mathrm{diag}(a_1,\cdots , a_n)$ denotes
a diagonal matrix whose diagonal entries are $a_1,\cdots, a_n$. Vectorization of matrix $\mathbf{A}$ is represented by $\mathrm{vec}(\mathbf{A})$.
The eigenvector corresponding to the largest eigenvalue of matrix $\mathbf{A}$ is denoted as $\boldsymbol{\lambda}_{\max}(\mathbf{A})$.
$\triangleq$ means ``defined as''.
Expectation and the real part of a complex number are denoted by $\mathbb{E}[\cdot]$ and $\Re(\cdot)$, respectively. 
The operation $\mathrm{Abs}(\mathbf{a})$ constructs a vector by extracting the magnitudes of the elements of vector $\mathbf{a}$, and $\mathrm{unt}(\mathbf{a})$ forms a vector whose elements are $\frac{a_1}{|a_1|},\cdots,\frac{a_n}{|a_n|}$.
The Hadamard product between two matrices is denoted by $\circ$. 

\section{System Model}
\begin{figure}
	\centering\includegraphics[width=6.4cm]{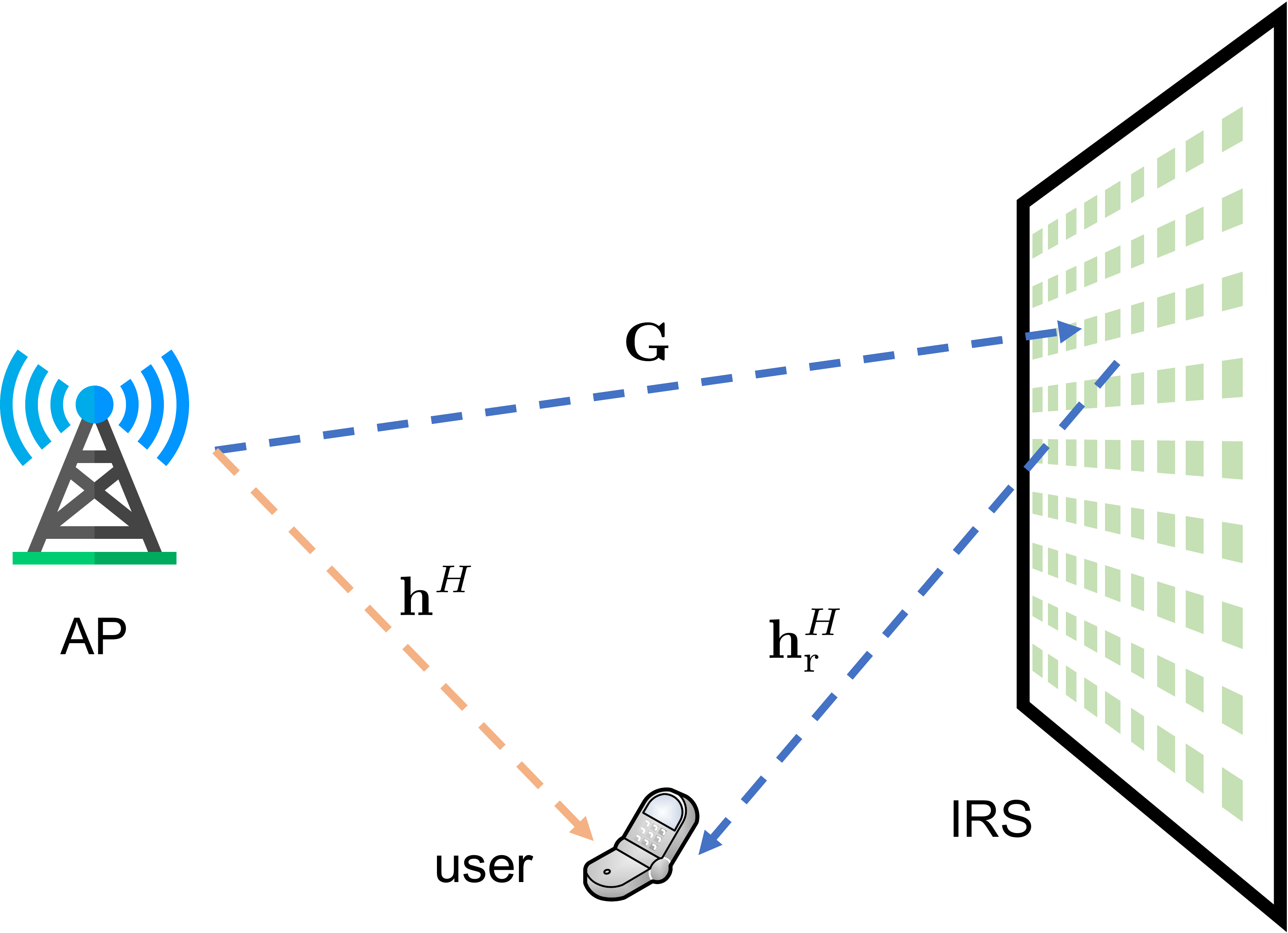}
	\caption{An IRS-assisted point-to-point MISO wireless communication system.}\label{model}
\end{figure}
Consider a single-user MISO wireless communication system, as shown in Fig. \ref{model}, where an $\Nt$-antenna AP transmits signals to a single-antenna user with the help of a passive IRS that employs $M$ phase shifters. 
The phase shifters are configurable and programmable via an IRS controller. Furthermore, we assume a quasi-static flat-fading channel model and perfect channel state information (CSI) knowledge at both the AP and the IRS\footnote{In practice, CSI can be accurately and efficiently obtained via various channel estimation techniques. 
The results in this paper serve as theoretical performance upper bounds
for the considered system, and  provide guidelines for the system design  when the CSI is
not perfectly known.}.
The received baseband signals at the user can be written as
\begin{equation}
y = \left(\mathbf{h}_\mathrm{r}^H\mathbf{\Phi}\mathbf{G}+\mathbf{h}^H\right)\mathbf{f}x+n,
\end{equation}
where $\mathbf{h}_\mathrm{r}\in\mathbb{C}^{M\times1}$ denotes the reflecting channel vector from the IRS to the user.
The phase shift matrix $\mathbf{\Phi}$  of the IRS is denoted by $\mathbf{\Phi}=\mathrm{diag}(e^{\jmath \theta_1},e^{\jmath \theta_2},\cdots,e^{\jmath \theta_M})$, where $\theta_i$ denotes the phase shift of the $i$-th reflecting element of the IRS\footnote{In general, the main diagonal elements of matrix $\mathbf{\Phi}$ may take the form $\beta e^{\jmath \theta_k}$ when the reflection loss $\beta\in[0,1]$ is considered. However, the value of $\beta$ does not affect the algorithm design. Therefore, in this paper, we set $\beta=1$ without loss of generality.} \cite{8647620}.
The channel matrix from the AP to the IRS is denoted by $\mathbf{G}\in\mathbb{C}^{M\times\Nt}$, and the linear beamforming vector at the AP is denoted by $\mathbf{f}\in\mathbb{C}^{\Nt\times1}$.
The direct channel vector between the AP and the user is represented by $\mathbf{h}\in\mathbb{C}^{\Nt\times1}$.
The transmitted signal is denoted by $x$, where $\mathbb{E}[|x|^2]=1$ without loss of generality. 
$n$ is additive complex Gaussian noise with variance $\sigma^2$.

In this paper, our goal is to maximize the achievable spectral efficiency by optimizing both the beamforming vector $\mathbf{f}$ and the phase shift matrix $\mathbf{\Phi}$. The spectral efficiency is given by
\begin{equation}
R=\log\left(1+\frac{\left|(\mathbf{h}_\mathrm{r}^H\mathbf{\Phi}\mathbf{G}+\mathbf{h}^H)\mathbf{f}\right|^2}{\sigma^2}\right),
\end{equation}
and the resulting optimization problem can be formulated as
\begin{equation}\label{eq3}
\begin{aligned}
&\underset{\mathbf{f},\mathbf{\Phi}}{\mathrm{maximize}} && \left|\left(\mathbf{h}_\mathrm{r}^H\mathbf{\Phi}\mathbf{G}+\mathbf{h}^H\right)\mathbf{f}\right|^2\\
&\mathrm{subject\thinspace to}&&\left\Vert\mathbf{f}\right\Vert^2\le P\\
&&&\mathbf{\Phi}=\mathrm{diag}\left(e^{\jmath \theta_1},e^{\jmath \theta_2},\cdots,e^{\jmath \theta_M}\right),
\end{aligned}
\end{equation}
where $P>0$ is the given total transmit power.

\emph{Remark 1:} The spectral efficiency can also be maximized with respect to the phase shifts $\{\theta_i\}_{i=1}^M$ at the IRS. The constraints are then convex and given by $0\le \theta_k\le 2\pi$, for $i\in\{1,2,\cdots,M\}$. However, the objective function is still non-convex with respect to the phase shifts $\{\theta_i\}_{i=1}^M$. Therefore, the optimization problem in \eqref{eq3} is basically a non-convex problem. To the best of the authors' knowledge, the globally optimal solution of non-convex optimization problems with unit modulus constraints is in general not tractable. In the next section, we shall propose two efficient algorithms that lead to locally optimal solutions.

\section{Design of IRS-Assisted MISO Wireless Systems}

\subsection{Problem Formulation}
Similar to the derivation steps in \cite[Eqs. (10) and (11)]{8647620}, the optimization problem in \eqref{eq3} can be reformulated as
\begin{equation}\label{eq4}\mathcal{P}_1:\quad
\begin{aligned}
&\underset{\mathbf{v}\in\mathbb{C}^{M+1}}{\mathrm{maximize}} && \mathbf{v}^H\mathbf{Rv}\\
&\mathrm{subject\thinspace to}&&|v_i|=1,\quad i\in\{1,2,\cdots,M+1\},
\end{aligned}
\end{equation}
where $\mathbf{v}=[\mathbf{x}^T,t]^T$, $\mathbf{x}=\left[e^{\jmath \theta_1},e^{\jmath \theta_2},\cdots,e^{\jmath \theta_M}\right]^H$, $t\in\mathbb{R}$, and
\begin{equation}\label{eq5}
\mathbf{R}=\begin{bmatrix}
\mathrm{diag}\left(\mathbf{h}_\mathrm{r}^H\right)\mathbf{GG}^H\mathrm{diag}\left(\mathbf{h}_\mathrm{r}\right)&
\mathrm{diag}\left(\mathbf{h}_\mathrm{r}^H\right)\mathbf{G}\mathbf{h}\\
\mathbf{h}^H\mathbf{G}^H\mathrm{diag}\left(\mathbf{h}_\mathrm{r}\right)&0
\end{bmatrix}.
\end{equation}
Note that optimization variable $\mathbf{v}$ in $\mathcal{P}_1$ is composed of an auxiliary variable $t$ and the phase shifts $\{\theta_i\}_{i=1}^M$. Once the solution for $\mathbf{v}$ in $\mathcal{P}_1$ is obtained, the corresponding beamforming vector $\mathbf{f}$ is optimally given by the maximum ratio transmission (MRT) strategy, i.e., 
\begin{equation}\label{eq6}
\mathbf{f}=\sqrt{P}\frac{\mathbf{G}^H\mathrm{diag}\left(\mathbf{h}_\mathrm{r}\right)\mathbf{x}+\mathbf{h}}{\left\Vert\mathbf{G}^H\mathrm{diag}\left(\mathbf{h}_\mathrm{r}\right)\mathbf{x}+\mathbf{h}\right\Vert}.
\end{equation}

\emph{Remark 2:}  $\mathcal{P}_1$ is a quadratically constrained quadratic program (QCQP) with a concave objective function. Furthermore, different from the typical semidefinite constraints, the unit modus constraints $|v_i|=1$ are intrinsically non-convex. Therefore, $\mathcal{P}_1$ is NP-hard in the problem size $M+1$ \cite{pardalos1991quadratic}.

\emph{Remark 3:} Optimization problem $\mathcal{P}_1$ was solved with an SDR approach in \cite{8647620,wu2018intelligent}. In particular, an auxiliary optimization variable $\mathbf{V}=\mathbf{vv}^H$ was introduced to reformulate $\mathcal{P}_1$ as a semidefinite programming (SDP) problem with an additional rank-one constraint. By dropping the rank-one constraint and solving the SDP problem via standard convex optimization tools, the optimal solution for $\mathbf{V}$ can be obtained. However, there is no guarantee that the obtained solution $\mathbf{V}$ is a rank-one matrix. A Gaussian randomization approach was adopted \cite{8647620,wu2018intelligent}, which ensures that the value of the objective function is asymptotically at least $\pi/4$ of the optimal value  \cite{zhang2006complex}.
Nevertheless, the SDR approach can only provide an approximate solution for $\mathbf{v}$.
In addition, solving an SDP problem is computationally expensive for
large IRS sizes $M$, as will be discussed in detail later. In this paper, we take this state-of-the-art SDR approach as the baseline for solving $\mathcal{P}_1$, and propose two novel algorithms that achieve a better performance in terms of both spectral efficiency and computational complexity.

\subsection{Fixed Point Iteration}

In this subsection, we propose a fixed point iteration method to solve $\mathcal{P}_1$, as presented in the following lemma.
\begin{lem}
	A limit point of the following fixed point iteration is a locally optimal solution of $\mathcal{P}_1$:
	\begin{equation}\label{eq7}
	\mathbf{v}^{(t+1)}=\mathrm{unt}\left(\mathbf{Rv}^{(t)}\right).
	\end{equation}
\end{lem}
\begin{IEEEproof}
	First, we prove that the iteration in \eqref{eq7} converges in value $\left\Vert\mathbf{Rv}\right\Vert_1$. 
	In particular, the value of  $\left\Vert\mathbf{Rv}\right\Vert_1$ monotonically increases because
	\begin{equation}
	\begin{split}
		\left\Vert\mathbf{Rv}^{(t+1)}\right\Vert_1&=\underset{|v_i|=1}{\max}\quad\Re\left(\mathbf{v}^H\mathbf{Rv}^{(t+1)}\right)\\
		&\ge\Re\left(\left(\mathbf{v}^{(t)}\right)^H\mathbf{Rv}^{(t+1)}\right)\\
		&=\Re\left(\left(\mathbf{v}^{(t)}\right)^H\mathbf{R}\times\mathrm{unt}\left(\mathbf{Rv}^{(t)}\right)\right)\\
		&\overset{(a)}{=}\left\Vert\mathbf{Rv}^{(t)}\right\Vert_1,
	\end{split}
	\end{equation}
	where $(a)$ exploits $\mathbf{R}=\mathbf{R}^H$.
	In addition, we show that  the value of $\left\Vert\mathbf{Rv}\right\Vert_1$ is upper bounded by 
	\begin{equation}\label{eq9}
	\begin{split}
	\left\Vert\mathbf{Rv}\right\Vert_1&=\sum_{i=1}^{M+1}\left|\sum_{j=1}^{M+1}r_{i,j}v_j\right|
	\le\sum_{i=1}^{M+1}\sum_{j=1}^{M+1}\left|r_{i,j}v_j\right|\\
	&\overset{(b)}{=}\sum_{i=1}^{M+1}\sum_{j=1}^{M+1}\left|r_{i,j}\right|,\\
	\end{split}
	\end{equation}
	where $(b)$ exploits $|v_j|=1$, and the final term in \eqref{eq9} is equal to the constant term $\left\Vert\mathrm{vec}(\mathbf{R})\right\Vert_1$. Therefore, the iteration in \eqref{eq7} converges in value $\left\Vert\mathbf{Rv}\right\Vert_1$.
	
	According to \eqref{eq7}, a limit point $\bar{\mathbf{v}}$ of  the fixed point iteration can be characterized by
	\begin{equation}
	\mathbf{R}\bar{\mathbf{v}}=\mathrm{Abs}\left(\mathbf{R}\bar{\mathbf{v}}\right)\circ\bar{\mathbf{v}}.
	\end{equation}
	Since $\mathrm{Abs}\left(\mathbf{R}\bar{\mathbf{v}}\right)$ is real-valued and non-negative, a limit point of the iteration in \eqref{eq7} is a locally optimal solution for $\mathcal{P}_1$ according to \cite[Appendix B]{6698378}, which completes the proof.
\end{IEEEproof}

As shown in Lemma 1, a locally optimal solution of $\mathcal{P}_1$ can be found with a fixed point iteration, and the corresponding algorithm is presented in \textbf{Algorithm 1}, where $\epsilon>0$ is a small threshold for the increment of the objective value until convergence.
\begin{algorithm}[t]
	\caption{Fixed Point Iteration}
	\begin{algorithmic}[1]
		\STATE Construct an initial $\mathbf{v}^{(0)}$ and set $t=0$;
		\REPEAT 
		\STATE Perform the iteration according to \eqref{eq7};
		\STATE $t\leftarrow t+1$;
		\UNTIL $\left\Vert\mathbf{Rv}^{(t+1)}\right\Vert_1-\left\Vert\mathbf{Rv}^{(t)}\right\Vert_1\le\epsilon$;
		\STATE Take the first $M$ elements of $\left(\mathbf{v}^{(t+1)}\right)^*$ as the main diagonal elements of matrix $\mathbf{\Phi}$;
		\STATE Design the  beamformer at the AP according to \eqref{eq6}.
	\end{algorithmic}
\end{algorithm}

\begin{figure*}
	\centering
	\subfigure[Tangent space and Riemannian gradient]
	{
		\centering\includegraphics[height=4cm]{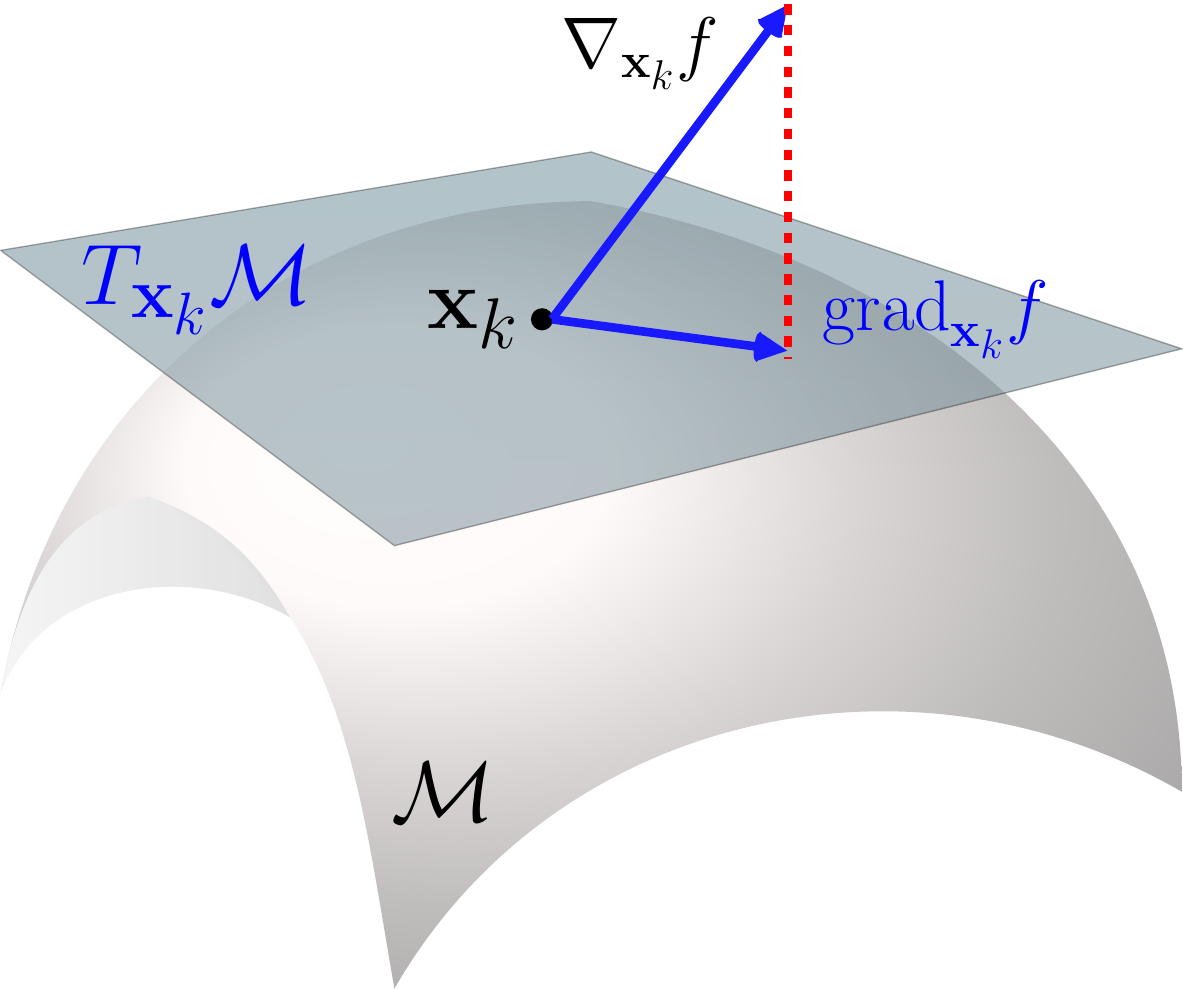}\label{fig21}
	}\quad\quad
	\subfigure[Vector transport]
	{
		\centering\includegraphics[height=4cm]{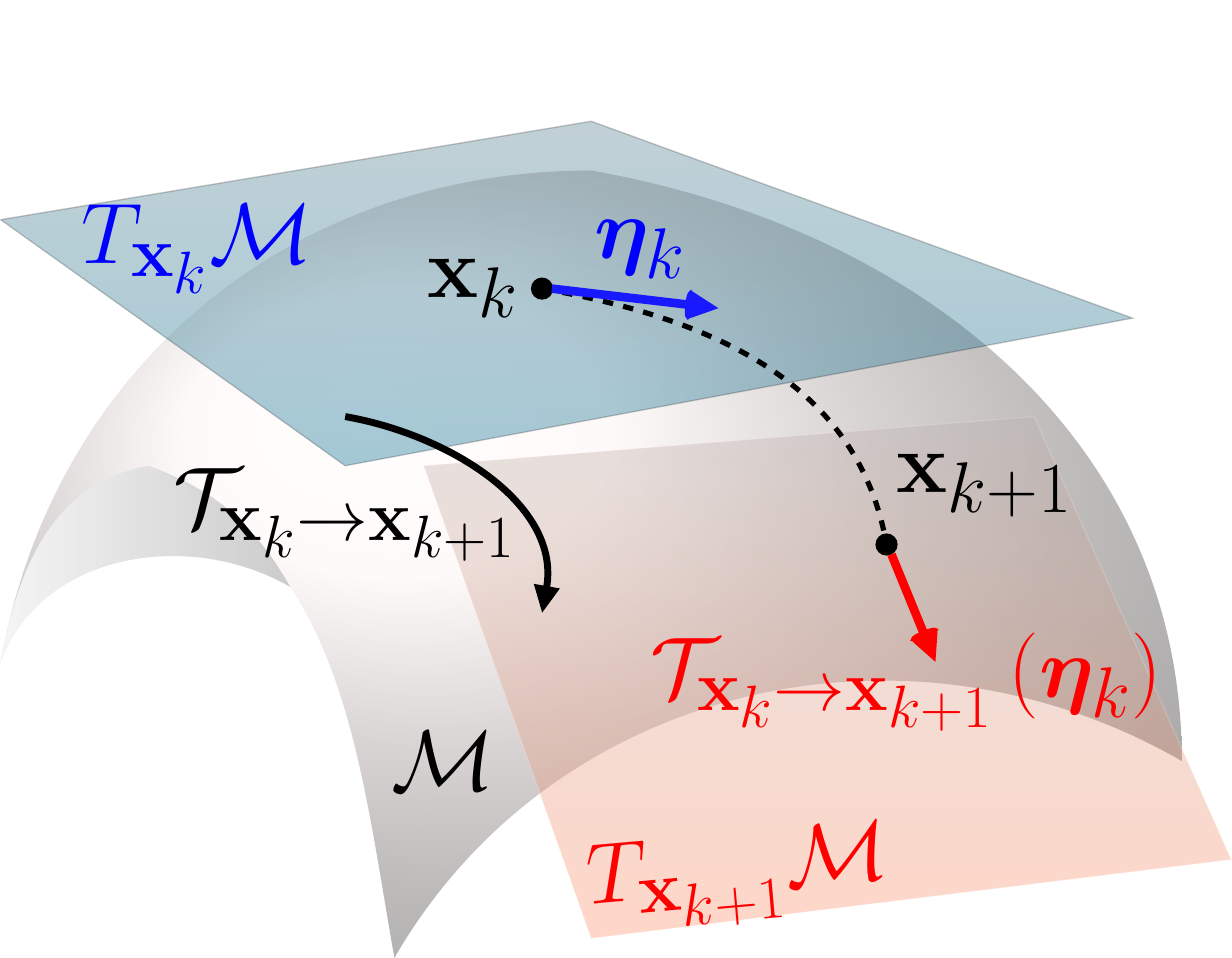}\label{fig22}
	}\quad\quad
		\subfigure[Retraction]
	{
		\centering\includegraphics[height=4cm]{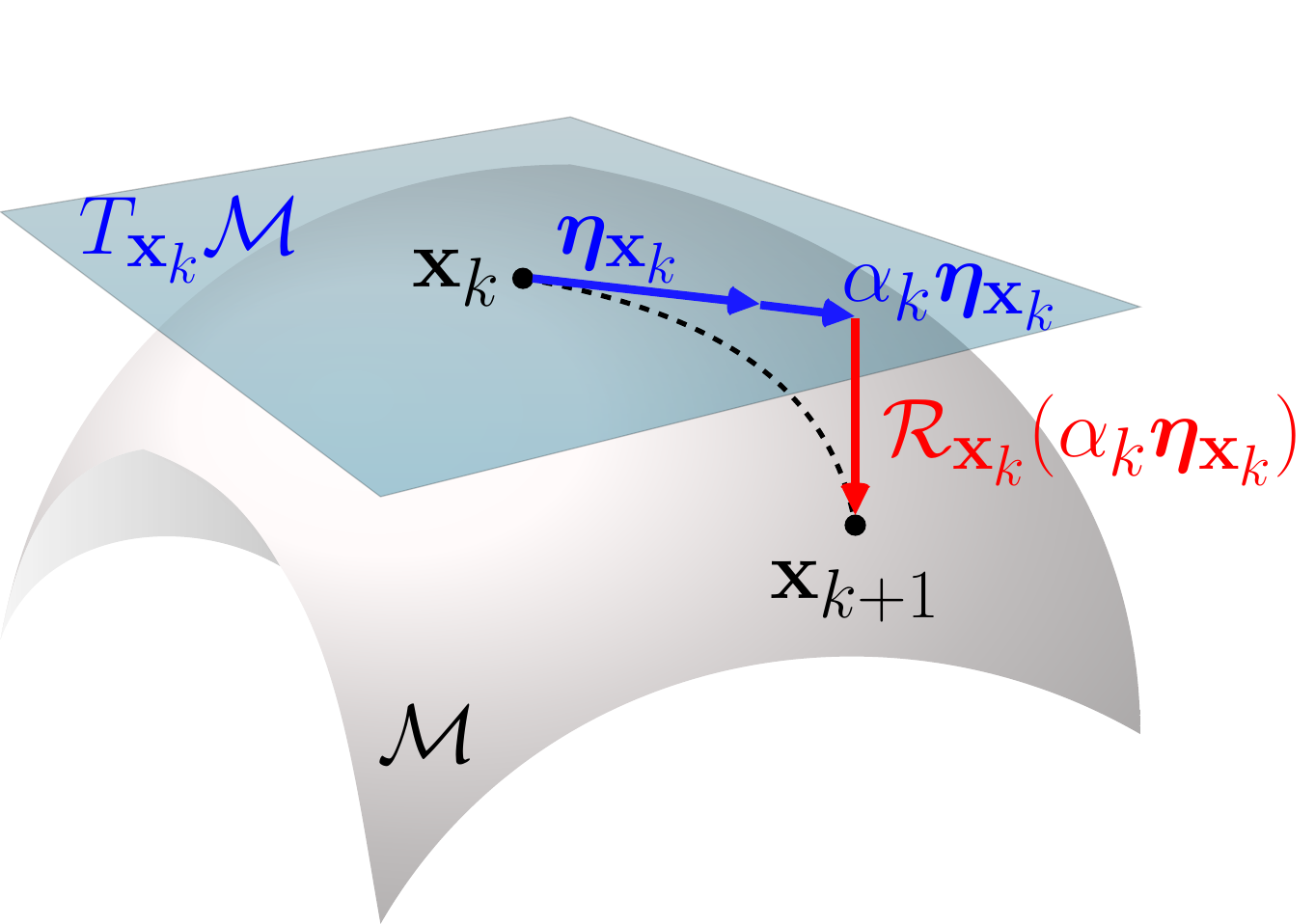}\label{fig23}
	}
	\caption{An illustrative example of the key steps in manifold optimization.}\label{manifold}
\vspace{-0.5em}
\end{figure*}
\subsection{Manifold Optimization}
Besides the fixed point iteration method, we propose a second algorithm that also finds a locally optimal solution for $\mathcal{P}_1$.
As pointed out in Remark 2, the unit modulus constraints $|v_i|=1$ are the main obstacles to solving $\mathcal{P}_1$. In this subsection, we show that $\mathcal{P}_1$ can be efficiently solved by manifold optimization, as the unit modulus constraints define a Riemannian manifold. 

In particular, the rich geometry of Riemannian manifolds makes it possible
to define gradients of cost functions. More importantly, the optimization over a manifold is locally analogous to that in the Euclidean space. Therefore, optimization techniques that were developed for the Euclidean space, e.g., gradient descent and trust-region methods, have counterparts on manifolds. There are some recent applications of manifold optimization in wireless communications \cite{7397861,8674787}. In the following, we briefly introduce the main idea of manifold optimization, see also Fig. \ref{manifold}.   

We rewrite $\mathcal{P}_1$ as the following optimization problem
\begin{equation}\mathcal{P}_2:\quad
\begin{aligned}
&\underset{\mathbf{x}\in\mathbb{C}^{M}}{\mathrm{minimize}} && 
f(\mathbf{\mathbf{x}})=-\mathbf{x}^H\mathbf{A}\mathbf{x}-\mathbf{x}^H\mathbf{b}-\mathbf{b}^H\mathbf{x}\\
&\mathrm{subject\thinspace to}&&|x_i|=1,\quad i\in\{1,2,\cdots,M\},
\end{aligned}
\end{equation}
where
\begin{equation}
\begin{split}
\mathbf{A}&=\mathrm{diag}\left(\mathbf{h}_\mathrm{r}^H\right)\mathbf{GG}^H\mathrm{diag}\left(\mathbf{h}_\mathrm{r}\right),\\
\mathbf{b}&=\mathrm{diag}\left(\mathbf{h}_\mathrm{r}^H\right)\mathbf{G}\mathbf{h}.\\
\end{split}
\end{equation}
The  unit modulus constraints $|x_i|=1$ form a complex circle manifold $\mathcal{M}=\{{\mathbf{x}\in\mathbb{C}^M:|{x}_1|=|{x}_2|=\cdots=|{x}_M|=1}\}$. Therefore, the search space of $\mathcal{P}_2$ is the product of $M$ circles in the complex plane, which is a Riemannian submanifold of $\mathbb{C}^M$ with the product geometry. More background on manifolds  can be found in \cite{absil2009optimization}.

For any point $\mathbf{x}_k$ on a manifold, the \emph{tangent space} is composed of all the tangent vectors that tangentially pass through $\mathbf{x}_k$. In particular, each tangent vector represents one direction along which one can move from $\mathbf{x}_k$ to optimize the objective function. For the complex circle manifold $\mathcal{M}$, the tangent space at $\mathbf{x}_k$ is given by

\begin{equation}
T_{\mathbf{x}_k}\mathcal{M}=\left\{\mathbf{z}\in\mathbb{C}^M:\Re\left\{\mathbf{z}\circ\mathbf{x}_k^*\right\}=\mathbf{0}_M\right\}.
\end{equation}

Similar to the Euclidean space, there is one tangent vector (direction) with the steepest increase of the objective function, called the \emph{Riemannian gradient}. As the complex circle manifold $\mathcal{M}$ is a Riemannian submanifold of $\mathbb{C}^M$, the {Riemannian gradient} of a function $f$ at point $\mathbf{x}_k$, denoted by $\mathrm{grad}_{\mathbf{x}_k}f$, is  the orthogonal projection of the Euclidean gradient $\nabla_{\mathbf{x}_k}f$ onto the tangent space  $T_{\mathbf{x}_k}\mathcal{M}$, as shown in Fig. \ref{fig21}. Therefore, the Riemannian gradient at a point $\mathbf{x}_k$ on the complex circle manifold $\mathcal{M}$ is given  by
\begin{equation}\label{rgradient}
\begin{split}
\mathrm{grad}_{\mathbf{x}_k}f=\nabla_{\mathbf{x}_k} f-\Re\{\nabla_{\mathbf{x}_k} f\circ \mathbf{x}_k^*\}\circ\mathbf{x}_k,
\end{split}
\end{equation}
where the Euclidean gradient of the objective function in $\mathcal{P}_2$ is given by
\begin{equation}\label{egradient}
\nabla_{\mathbf{x}_k} f=-2\left(\mathbf{Ax}_k+\mathbf{b}\right).
\end{equation}

With the Riemannian gradient $\mathrm{grad}_{\mathbf{x}_k}f$ at hand, 
abundant optimization techniques developed for the Euclidean space can be transplanted to manifold optimization. 
For instance, the update rule of the search direction of the  conjugate gradient (CG) method in the Euclidean space is given by
\begin{equation}\label{eq15}
\boldsymbol{\eta}_{k+1}=-\nabla_{\mathbf{x}_{k+1}}f+\beta_k\boldsymbol{\eta}_k,
\end{equation}
where $\boldsymbol{\eta}_k$ is the search direction at $\mathbf{x}_k$, and $\beta_k$ is chosen as the Polak-Ribiere parameter \cite{absil2009optimization}. However, $\boldsymbol{\eta}_k$ and $\boldsymbol{\eta}_{k+1}$ in manifold optimization lie in two different tangent spaces $T_{\mathbf{x}_k}\mathcal{M}$ and $T_{\mathbf{x}_{k+1}}\mathcal{M}$. Therefore, operations such as the sum in \eqref{eq15} that involve
different tangent spaces cannot be directly performed. To overcome this problem, a \emph{transport}, defined as the mapping of a tangent vector from one tangent space to another tangent space, is needed, as shown in Fig. \ref{fig22}. The vector transport for manifold $\mathcal{M}$ is given by
\begin{equation}\label{transport}
\begin{split}
\mathcal{T}_{\mathbf{x}_k\to\mathbf{x}_{k+1}}\left(\boldsymbol{\eta}_k\right)\triangleq T_{\mathbf{x}_k}\mathcal{M}&\mapsto T_{\mathbf{x}_{k+1}}\mathcal{M}:\\
\boldsymbol{\eta}_k&\mapsto \boldsymbol{\eta}_k-\Re\{\boldsymbol{\eta}_k\circ \mathbf{x}_{k+1}^*\}\circ\mathbf{x}_{k+1}.
\end{split}
\end{equation}
Analogous to \eqref{eq15}, the update rule for the search direction for the CG method on manifolds is given by
\begin{equation}\label{eq17}
\boldsymbol{\eta}_{k+1}=-\mathrm{grad}_{\mathbf{x}_{k+1}}f+\beta_k\mathcal{T}_{\mathbf{x}_k\to\mathbf{x}_{k+1}}\left(\boldsymbol{\eta}_k\right).
\end{equation}

\begin{algorithm}[t]

	\caption{CG Method Based Manifold Optimization}
	\begin{algorithmic}[1]
		\STATE Construct an initial $\mathbf{x}^{(0)}$ and set $k=0$;
		\STATE Calculate $\boldsymbol{\eta}_k=-\mathrm{grad}_{\mathbf{x}_k}f$ according to \eqref{rgradient};
		\REPEAT
		\STATE Choose Armijo backtracking line search step size $\alpha_k$ according to \cite[Eq. (59)]{shewchuk1994introduction};
		\STATE Find the next point $\mathbf{x}_{k+1}$ using retraction in \eqref{retraction}: 
		$\mathbf{x}_{k+1}=\mathcal{R}_{\mathbf{x}_k}(\alpha_k\boldsymbol{\eta}_k);$
		\STATE Determine Riemannian gradient $\mathrm{grad}_{\mathbf{x}_{k+1}}f$ according to \eqref{rgradient};
		\STATE Calculate the vector transport $\mathcal{T}_{\mathbf{x}_k\to\mathbf{x}_{k+1}}\left(\boldsymbol{\eta}_k\right)$ according to \eqref{transport};
		\STATE Choose Polak-Ribiere parameter $\beta_{k}$ \cite[p. 42]{shewchuk1994introduction};\label{s7}
		\STATE Compute conjugate search direction $\boldsymbol{\eta}_{k+1}$ with \eqref{eq17};
		\STATE $k\leftarrow k+1$;
		\UNTIL $\left\Vert\mathrm{grad}_{\mathbf{x}_k}f\right\Vert_2\le\epsilon$;
		\STATE Take $\mathbf{x}_{k+1}^*$ as the main diagonal elements of matrix $\mathbf{\Phi}$;
		\STATE Design the  beamformer at the AP according to \eqref{eq6}.
	\end{algorithmic}
\end{algorithm}

After determining the search direction $\boldsymbol{\eta}_k$ at $\mathbf{x}_k$, an operation called \emph{retraction} is used to find the destination on the manifold, as shown in Fig. \ref{fig23}. Specifically, retraction is a mapping from the tangent space to the manifold itself. For a point $\mathbf{x}_k$ on manifold $\mathcal{M}$, the retraction for the search direction $\boldsymbol{\eta}_k$ and step size $\alpha_k$ is given by
\begin{equation}\label{retraction}
\mathcal{R}_{\mathbf{x}_k}\left(\alpha_k\boldsymbol{\eta}_k\right)\triangleq T_{\mathbf{x}_k}\mathcal{M}\mapsto\mathcal{M}:
\alpha_k\boldsymbol{\eta}_k\mapsto\mathrm{unt}\left(\alpha_k\boldsymbol{\eta}_k\right).
\end{equation}

Now, the key steps used in each iteration of the manifold optimization have been introduced. The resulting algorithm for solving $\mathcal{P}_2$ is summarized in \textbf{Algorithm 2}. \textbf{Algorithm 2} is guaranteed to converge to a critical point of $\mathcal{P}_2$, i.e., the point where the Riemannian gradient of the objective function is zero \cite{absil2009optimization}.

\subsection{Discussion}

In this subsection, we provide some discussions on the initial points and the computational complexities of the proposed algorithms.

\emph{(1) Initialization:} For both \textbf{Algorithms 1} and \textbf{2}, an initialization for the phase shifts at the IRS is required. As both algorithms lead to locally optimal solutions, it is desirable to initialize the optimization variables with values that are ``close''  to the optimal solution. In this paper, we resort to a heuristic approach to construct the initial point. In particular, we relax the unit modulus constraints in $\mathcal{P}_1$ to a norm constraint $\left\Vert\mathbf{v}\right\Vert_2^2=M+1$, such that the resulting problem reduces to an eigenvalue problem whose optimal solution is given by
\begin{equation}
\tilde{\mathbf{v}}^\star=\sqrt{M+1}\boldsymbol{\lambda}_{\max}\left(\mathbf{R}\right).
\end{equation}
Then, we perform a phase extraction of this solution to form a unit modulus vector as the initialization for \textbf{\textbf{Algorithm 1}}, i.e.,
\begin{equation}
\mathbf{v}^{(0)}=\mathrm{unt}\left(\tilde{\mathbf{v}}^\star\right),
\end{equation}
and the first $M$ elements of $\mathbf{v}^{(0)}$ are selected as the initial point $\mathbf{x}^{(0)}$ for \textbf{Algorithm 2}.

\emph{(2) Computational complexity:} In the fixed point iteration method, a closed-form solution \eqref{eq7} is given in each iteration of \textbf{Algorithm 1}. In addition, the worst-case computational complexity of the CG method in \textbf{Algorithm 2} is $\mathcal{O}\left(M^{1.5}\right)$ \cite{shewchuk1994introduction}. In contrast, the state-of-the-art SDR approach in \cite{8647620,wu2018intelligent} entails a computational complexity of $\mathcal{O}\left((M+1)^{6}\right)$, which is prohibitively high compared to the proposed algorithms\footnote{The computational complexities of \textbf{Algorithm 2} and the SDR method apply to the entire algorithm not just to one iteration.}. 
The complexities of the proposed algorithms will also be compared via simulation in the next section.

\section{Simulation Results}
In this section, we evaluate the performance of the proposed algorithms by using the SDR method as benchmark. All channels are assumed to be
independent Rayleigh fading, and the path loss exponent is set to 3 with reference distance $10$ m. The total transmit power is $P=5$ dBm while the noise power at the user is set to $\sigma^2=-80$ dBm. All simulation results in this section are averaged over 1000 channel realizations. The stopping criterion for
convergence for both proposed algorithms is that the increment of the objective
function is less than $\epsilon=10^{-6}$.

\subsection{Average Spectral Efficiency vs. AP-User Distance}
\begin{figure}[t]
	\centering\includegraphics[width=6.5cm]{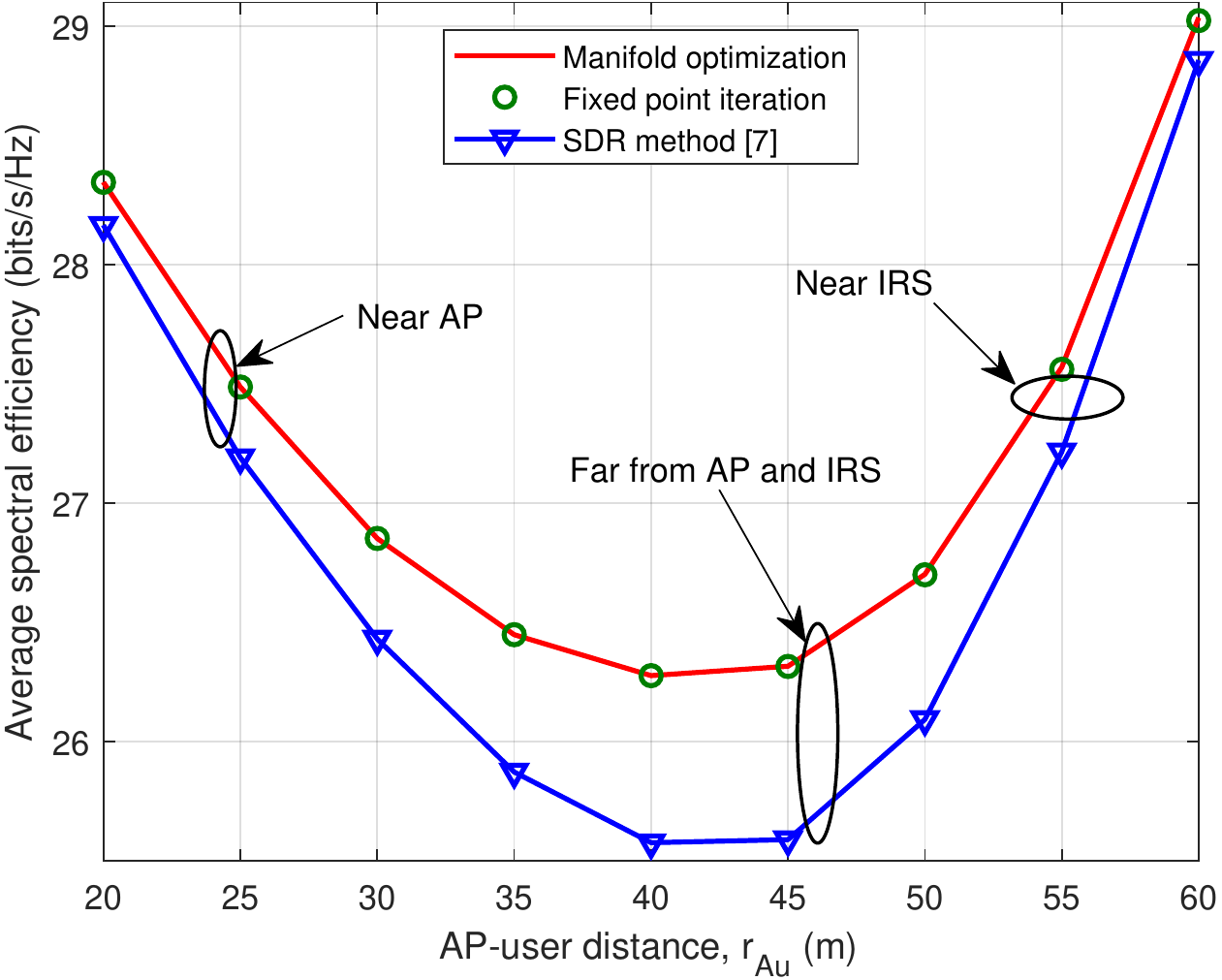}
	\caption{Spectral efficiency achieved by different algorithms when $r_\mathrm{AI}=50$ m, $r_\mathrm{Au}+r_\mathrm{Iu}=70$ m, $\Nt=8$, and $M=10$.}\label{fig1}
\end{figure}
We consider a system where the distance $r_\mathrm{AI}$ between the AP and IRS  is fixed to $50$ m. To evaluate the impact of the user position, we assume that the user moves such that the sum  of the AP-user distance $r_\mathrm{Au}$ and the IRS-user distance $r_\mathrm{Iu}$ is $r_\mathrm{Au}+r_\mathrm{Iu}=70$ m. In Fig. \ref{fig1}, we compare the spectral efficiency achieved for different AP-user distances. We first observe that the proposed manifold optimization-based algorithm and fixed point iteration method achieve almost the same spectral efficiency, and both significantly outperform the SDR method. This is mainly because our proposed algorithms are guaranteed to find locally optimal solutions of $\mathcal{P}_1$ and $\mathcal{P}_2$, respectively, while the SDR method yields only an approximate solution as mentioned in Remark 3.

When the user moves close to the AP, it  benefits from the strong direct channel $\mathbf{h}$ between the AP and the user, and therefore achieves a high spectral efficiency. A similar phenomenon can also be observed when the user moves close to the IRS, such that the user benefits from a strong reflecting channel $\mathbf{h}_\mathrm{r}$. Under these two conditions, the optimal joint design of the beamformer at the AP and the IRS phase shifts is less important. Hence, the performance gap between the proposed algorithms and the SDR method is relatively small in these two regimes in Fig. \ref{fig1}. In contrast, when the user is located relatively far from both the AP and the IRS, e.g., $r_\mathrm{Au}=40$ m, the optimal joint design of the beamformer at the AP and the IRS phase shifts is needed to achieve a satisfactory communication performance. In this case, our proposed algorithms are good candidates for improving the spectral efficiency.

\subsection{Computational Complexity}
\begin{figure}[t]
	\centering\includegraphics[width=6.5cm]{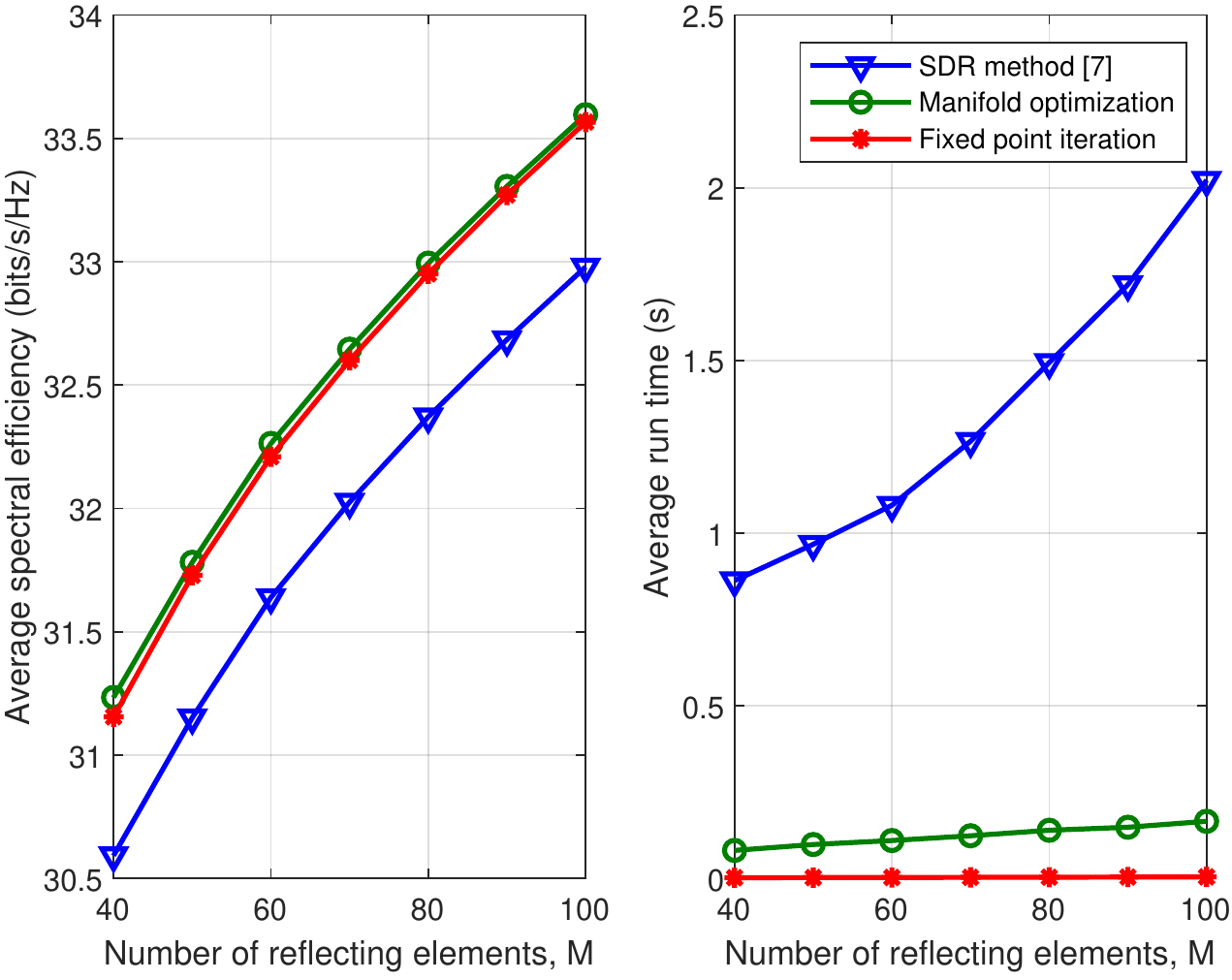}
	\caption{Average spectral efficiency and run time for different numbers of IRS reflecting elements when $\Nt=5$, $r_\mathrm{AI}=r_\mathrm{Au}=60$ m, and $r_\mathrm{Iu}=10$ m.}\label{fig2}
\end{figure}

In Fig. \ref{fig2} (left hand side), the average spectral efficiency achieved for different numbers of reflecting elements, $M$, is plotted. As can be observed, there is a small gap between the two proposed algorithms for large IRS sizes $M$. Note that both of the proposed algorithms lead to locally optimal solutions. As the number of non-convex constraints defining  the feasible sets of $\mathcal{P}_1$ and $\mathcal{P}_2$ increases when the IRS size $M$ grows large, manifold optimization is more likely to escape saddle points of the resulting non-convex problem compared to the fixed point approach, which results in a (slightly) higher spectral efficiency. 
In addition, in Fig. \ref{fig2} (right hand side), we plot the average run time of the proposed algorithms. As can be observed, the proposed algorithms require much less run time than the SDR method, which confirms the discussion at the end of Section III-D. 
In summary, the proposed algorithms are not only spectral-efficient but also computationally-efficient.

\subsection{Beyond Massive MIMO}
\begin{figure}[t]
	\centering\includegraphics[width=6.5cm]{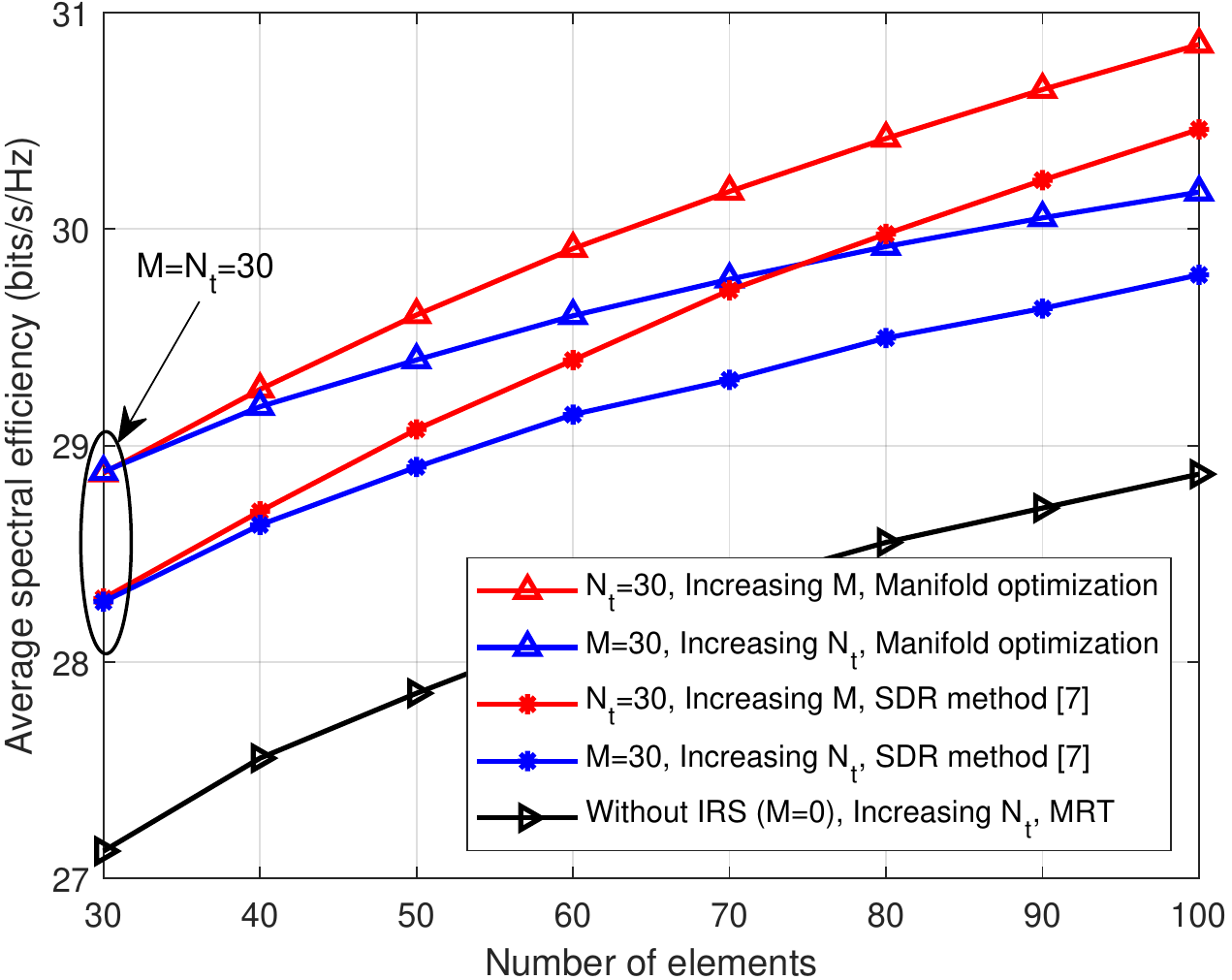}
	\caption{Average spectral efficiency achieved for different values of $M$ and $\Nt$  when  $r_\mathrm{AI}=50$ m, $r_\mathrm{Au}=40$ m, and $r_\mathrm{Iu}=30$ m.}\label{fig3}
\end{figure}

In conventional cellular networks without IRSs, deploying large-scale antenna arrays is an effective way to boost the network capacity. The black curve in Fig. \ref{fig3} illustrates this effect, where  optimal MRT beamforming is adopted to align the beam to the direct channel $\mathbf{h}$.
For the IRS-assisted system considered in this paper, the red curves in Fig. \ref{fig3} represent the spectral efficiency achieved for increasing values of $M$, while keeping the transmit antenna array size as $\Nt=30$. As the performance of the two proposed algorithms is almost the same, the curves for the fixed point iteration method are omitted for clarity of presentation. On the other hand, the blue curves in Fig. \ref{fig3} depict the spectral efficiency 
achieved for different numbers of transmit antenna elements when using an IRS
with $M=30$. We observe that both considered systems with IRSs significantly outperform the MRT strategy without IRSs, which confirms the effectiveness of incorporating IRSs into wireless communication systems.

More importantly, Fig. \ref{fig3} clearly shows that 
increasing the number of IRS reflecting elements is more
efficient than enlarging the transmit antenna array size in terms of spectral efficiency. It can also be observed that the performance gain increases with the number of elements. 
Furthermore, additional RF chains and power amplifiers need to be deployed for increasing the number of antenna elements, which leads to a more energy-consuming design compared to deploying large-scale  passive IRSs. Therefore, we conclude that IRS-assisted wireless systems are more spectral- and energy-efficient than conventional wireless systems.

\section{Conclusions}
This paper investigated the joint design of the beamformer at the AP and the  IRS phase shifts for an IRS-assisted wireless communication system. It was shown that the proposed algorithms, i.e., the fixed point iteration and manifold optimization methods, are effective in tackling the unit modulus constraints, which are the main obstacles for optimizing the IRS phase shifts. One particular contribution of this paper is the identification of the manifold structure of the IRS phase shifts, as this allows the application of  powerful manifold optimization techniques. Simulation results revealed that IRSs have substantial potential for the establishment of high-speed green communication networks.
%


\bibliographystyle{IEEEtran}
%
\bibliography{bare_conf}

\end{document}